\documentclass{article}

\usepackage{arxiv}

\usepackage[T1]{fontenc} 
\usepackage[utf8]{inputenc}

\usepackage{microtype}
\usepackage{todonotes}
\usepackage{graphicx}
\usepackage{subfigure}
\usepackage{booktabs}
\usepackage{amsmath}
\usepackage{float}
\usepackage{booktabs}
\usepackage{multirow}
\usepackage{microtype}
\usepackage{amsfonts} 
\usepackage{url}  

\usepackage{cleveref}
\usepackage{algorithmic}
\usepackage{algorithm}

\title{
    To compress or not to compress: \\
    Understanding the Interactions between
    \\Adversarial Attacks and Neural Network Compression}

\author{
Yiren Zhao* \\
Computer Laboratory \\
University of Cambridge \\
\And
Ilia Shumailov* \\
Computer Laboratory \\
University of Cambridge \\
\And
Robert Mullins \\
Computer Laboratory \\
University of Cambridge \\
\And
Ross Anderson \\
Computer Laboratory \\
University of Cambridge \\
}

\begin{document}
\maketitle

\begin{abstract}
As deep neural networks (DNNs) become widely used,
pruned and quantised models are becoming ubiquitous on edge devices;
such compressed DNNs lower the computational requirements.
Meanwhile, multiple recent studies show ways of constructing
adversarial samples that make DNNs misclassify.
We therefore investigate the extent to which
adversarial samples are transferable between uncompressed and compressed
DNNs.
We find that such samples remain transferable for both pruned and
quantised models.
For pruning,
adversarial samples at high sparsities are marginally less transferable.
For quantisation,
we find the transferability of adversarial samples is highly sensitive to
integer precision.
\end{abstract}

\section{Introduction}
Deep Neural Networks (DNNs)
perform well on a wide range of tasks, including
image classification \cite{krizhevsky2012imagenet},
object detection \cite{ren2015faster},
reading comprehension \cite{seo2016bidirectional} and
machine translation \cite{bahdanau2014neural}.
They have proved to be an efficient
method of harvesting information from large amounts of data
and are expected to be ubiquitous in the future.
Despite these successes, two
questions remain crucial for
deploying them in embedded systems.
First,
their substantial computational and memory
requirements can make deployment challenging
on power-limited devices.
Second,
as they start to appear in safety-critical applications,
their reliability and security become a serious issue.

In order to compute DNNs efficiently on embedded systems,
researchers have proposed various compression methods.
%The research has two main branches: 1) efforts to reduce their computation requirements; and 2) efforts to implement custom accelerators.
\textit{Pruning} directly reduces the number of parameters
of DNNs -- this reduction translates to
fewer data movements and thus saves energy directly.
\textit{Quantisation} is another popular compression technique
-- it simultaneously reduces the memory footprint and decreases the
energy cost of multiplications.
Both compression methods are widely deployed on DNN accelerators.
For instance, Efficient Inference Engines (EIE) use pruning, quantisation
and encoding techniques for energy efficiency \cite{Han2016EIE};
the Sparse CNN (SCNN) accelerator first requires network parameters
to be pruned and encoded, then performs
computations directly using
the encoded data format \cite{Parashar2017SCNN}.
It seems likely that pruning, quantisation and other compression
techniques will be used
for future DNNs on embedded devices, even where hardware accelerators are also used.

Over the past five years, research has found DNNs to be sensitive to small
perturbations of the input images, with the result that they can often be fooled
easily using specially-crafted adversarial inputs
\cite{DBLP:journals/corr/SzegedyZSBEGF13}.
Such adversarial samples are a real concern for
safety-critical systems; attackers might try to
manipulate autonomous vehicles \cite{eykholt2018robust}
or break into smart phones by tricking the
speaker recognition system \cite{197215}.

In this paper, we study the portability of adversarial samples. Might an attacker learn how to break
into widely-deployed low-cost systems and then use the same
adversarial samples as a springboard to attack other related systems?

We make the following contributions in this paper.

\begin{itemize}
    \item We investigated the effects of different
        DNN compression mechanisms on adversarial attacks.
    \item We have developed the first
        compression-aware machine learning attack taxonomy
        and used it to evaluate the transferability of
        adversarial samples between compressed and uncompressed models.
    \item As for pruning, we found that adversarial samples are transferable
    between compressed and uncompressed models.
    However, adversarial samples
    generated from uncompressed models are marginally
    less effective on compressed models at preferred sparsities, and
    adversarial samples generated from extremely sparse models are no longer
    effective on the baseline model.
    \item As for quantisation, we found that adversarial samples are transferable
    between compressed and uncompressed models.
    However, a reduction in integer precision
    provides clipping effects and marginally limits transferability
    in fast gradient based attacks.
\end{itemize}

\section{Related Work}
\subsection{Pruning}
Pruning directly reduces the number of parameters in a DNN model, 
and thus the number of 
off-chip to on-chip data transfers 
on modern DNN accelerators \cite{chen2016eyeriss}.
If the architecture allows, 
pruning may also reduce the computation cost \cite{kim2018zena}.
Consider a weight tensor ($\mathbf{W_n}$); 
fine-grained pruning is simply performing an 
element-wise multiplication ($\odot$) 
between a mask operator $\mathbf{M_n}$ and 
the original weight tensor ($\mathbf{W_n}$).

\begin{equation}
    \mathbf{W_n}' = \mathbf{W_n} \odot \mathbf{M_n}
  \label{equ:maskfunc1}
\end{equation}

Han et al. first proposed pruning a DNN 
by applying a threshold to the DNN's parameters \cite{han2015deep}.
In this case, the mask ($\mathbf{M_n}$) 
consists of thresholding by a single value $\alpha$.

\begin{equation}
    \mathbf{M_n} = h_k(\mathbf{W_n}^{(i,j)}) = \begin{cases}
        0 & \text{    if }\alpha > |\mathbf{W_k}^{(i,j)}| \\
        1 & \text{    otherwise} \\
    \end{cases}
  \label{equ:maskfunc2}
\end{equation}

Using this simple one-shot pruning technique, 
Han et al. were able to reduce the number of parameters 
in AlexNet by 9x and VGG16 by 13x \cite{han2015deep}. 
In their implementation, the masking and 
fine-tuning happen iteratively but the masked values 
are not allowed to recover in later stages.

Guo et al. subsequently proposed dynamic network surgery (DNS), 
which allows pruned parameters to recover at later stages~\cite{guo2016dynamic}.
The approach is to condition the mask 
using the following equation, 
where $\alpha$ and $\beta$ are two constants.

\begin{equation}
    \mathbf{M_n} = h_k(\mathbf{W_n}^{(i,j)}) = \begin{cases}
        0     & \text{    if }\alpha > |\mathbf{W_k}^{(i,j)}| \\
        \mathbf{M_n}^{(i,j)} & \text{    if }\alpha \leq |\mathbf{W_k}^{(i,j)}| \leq \beta\\
        1 & \text{    otherwise} \\
    \end{cases}
  \label{equ:maskfunc3}
\end{equation}

Values that become bigger at later stages
are allowed to rejoin the fine-tuning process.
Guo et al. demonstrated higher compression rates on a large range 
of networks compared to Han et al.
In this paper, we generate pruned DNNs using the DNS method.

\subsection{Quantisation}
Quantisation refers to using fewer bits for parameters in a DNN
than the standard 32-bit single-precision floating-point representation
used on modern CPUs and GPUs.
Hubara et al. showed that 
low-precision fixed-point numbers can be used for neural-network 
inference with nearly no loss of accuracy~\cite{HubaraCSEB16}.
In the extreme case, 
the parameters of a DNN can be quantised to 
either binary or ternary values~\cite{courbariaux2016binarized,li2016ternary}.
Such aggressive quantisation 
can greatly speed up DNN hardware accelerators 
but suffers from significant loss of accuracy.
For resource-constrained devices, 
a low-precision fixed-point representation 
can give a balance between accuracy 
and performance~\cite{lin2016fixed}.
The narrower bitwidth means direct reductions 
in memory requirement 
and fixed-point multiplications are less 
computationally expensive compared with standard 
single-precision floating point.
In this paper, 
we generate models that use fixed-point parameters 
at various levels of precision.

\subsection{Adversarial Attacks}

Szegedy et al. discovered that, 
despite generalising well, models trained on huge datasets are all vulnerable to adversarial samples~\cite{DBLP:journals/corr/SzegedyZSBEGF13}. 
Misclassification can even happen with imperceptible perturbations 
of the data samples.
All the samples they used were within the expected data distribution and 
only a small specially-crafted amount of noise was added. 
They observed that models of 
different configurations, trained on different datasets, 
misclassify the same samples. 
Finally, they noted that training a model on adversarial samples helps
make it more robust against them.
However, this defence is not always practical; their approach based on L-BFGS requires an expensive constrained optimisation with multiple iterations. 

A follow-up paper by Goodfellow et al. \cite{goodfellow2014explaining}
explored the underlying reasons 
for the existence and generalisability of adversarial samples.
They argue that such samples are an artefact of 
high-dimensional dot-products, and attacks are generalisable because
different models learn similar functions when trained to perform the 
same task. Additionally, they presented two methods to generate 
adversarial samples in a white-box setting, the
fast gradient method (FGM) and the fast gradient sign method (FGSM). 
Finally, they discovered that RBF-based networks 
are much more resistant to adversarial samples. 

Papernot et al. \cite{DBLP:journals/corr/PapernotMJFCS15} came up with
another way to generate adversarial samples.
They use the gradients of a network to 
construct saliency maps 
for the input to discover which input values are so sensitive that a change 
can drive misclassification. 
They showed 
that their method has the flexibility of being used in both 
supervised and unsupervised settings 
and is capable of generating samples with a user-given priority 
on particular properties of the inputs. 
Finally, they also observed that
adversarial attacks become harder when models have been trained
with adversarial samples.

There is now a growing corpus of research 
on the transferability of adversarial samples
\cite{DBLP:journals/corr/SzegedyZSBEGF13, 
goodfellow2014explaining, papernot2016transferability, tramer2017space}. 
Transferability refers to the ability of an adversarial sample 
to evade the correct classification on two different classifiers trained to 
perform a similar task. 

Goodfellow et al.~\cite{goodfellow2014explaining} 
and Warde-Farley \& Goodfellow~\cite{warde201611} empirically found 
that adversarial examples usually occur in large, continuous spatial 
regions. Tramer et al.~\cite{tramer2017space} found out that each of the models differs in 
the dimensionality of its subspaces. A higher number of dimensions increases the 
chance that the subspaces of different models intersect, leading to 
transferable samples. 

Transferable adversarial samples are a real hazard for model deployment, as they are `break-once, run-anywhere': attacks developed on a particular type of classifier can
potentially be deployed everywhere.
Papernot et al. \cite{DBLP:journals/corr/PapernotMGJCS16} in particular have 
shown that an adversary can sometimes perform attacks without any 
knowledge of a model's internal parameters -- it can be enough to approximate a 
model with another known model and build adversarial samples against that instead. 
\section{Methodology}

\subsection{Attack Taxonomy and Threat Model}

\begin{figure*}[t]
\centering
\begin{minipage}{.5\linewidth}
  \centering
  \includegraphics[width=0.8\linewidth]{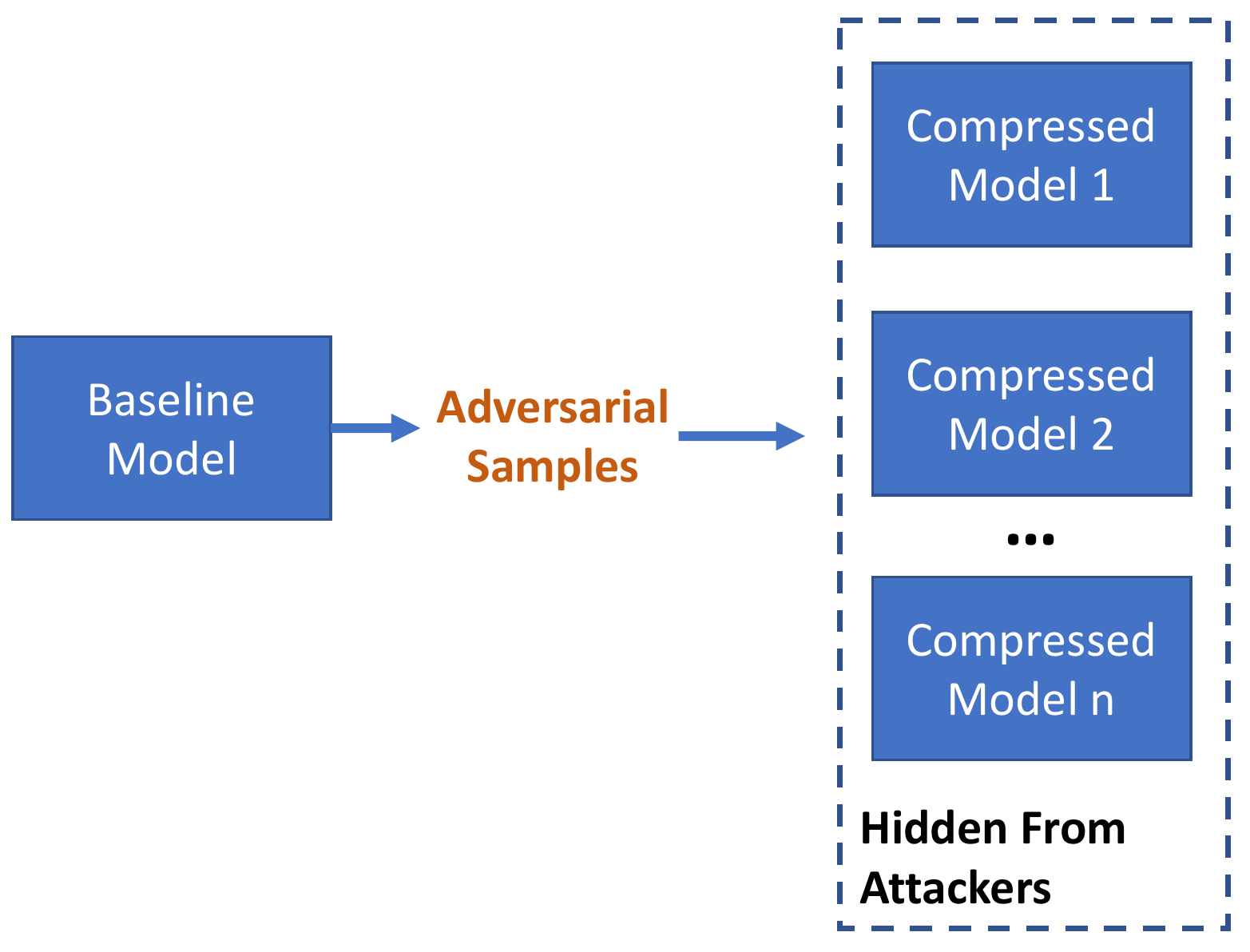}  
\end{minipage}%
\begin{minipage}{.5\linewidth}
  \centering
  \includegraphics[width=0.8\linewidth]{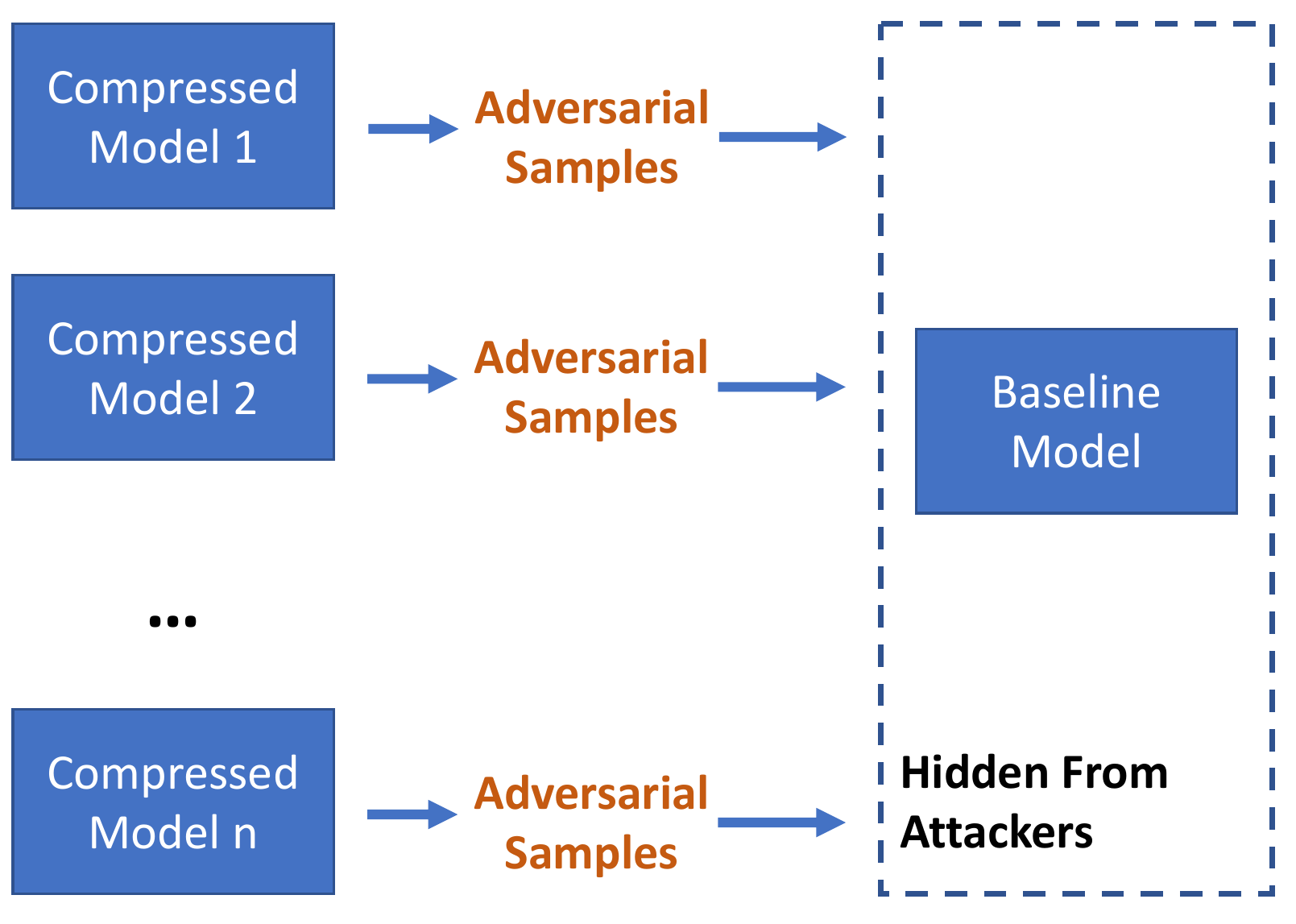}
\end{minipage}
\caption{Two different attack setups. Attackers only generate adversarial samples based on baseline model (left), or attackers generate adversarial samples based on compressed models (right).}
\label{fig:taxonomy}
\end{figure*}

In this paper, we are interested in
the interaction between adversarial attacks and model compression,
and we investigate three specific attack scenarios. 
By `compressed models' we will mean models that have pruned or quantised, while a `baseline model' means  
a pretrained network without any compression that is dense and whose parameters are 
represented using full-precision (float32) values.

\begin{itemize}
    \item Scenario 1: Adversarial attacks occur on each 
    individual compressed model, with 
    the adversarial examples generated and applied 
    on the same model.
    \item Scenario 2: Adversarial samples 
    are generated from the baseline model 
    but applied on each compressed model.
    \item Scenario 3: Adversarial samples are generated 
    from compressed models but applied on the baseline model.
\end{itemize}

In the first scenario, adversarial samples 
are generated from each compressed model. 
Attackers can access these compressed models fully, 
and generate adversarial samples for each one individually. This is 
the case where attackers buy products and figure out how to attack them.

The second scenario makes
the assumption that attackers can only access the baseline model
to generate adversarial samples,
which are then used to attack various compressed models. 
Attackers are not allowed to fetch any gradients from compressed models. 
This is the case where firms take publicly-available models 
and compress them to run more efficiently on edge devices. 
Attackers can find the public model and craft adversarial samples to attack derived proprietary devices.

The third scenario assumes that only compressed models are visible to attackers, 
and attackers generate adversarial samples 
using compressed models to attack the hidden baseline model. 
In practice, companies now deploy various compressed neural-network models 
on edge devices that are exposed to end-users. 
The assumption is the attackers can
access these models and create
adversarial samples from them to attack the hidden baseline model.
This then back leads to the second scenario; the attacker's knowledge
and toolkit can be transferred to other compressed products from the same firm.
\Cref{fig:taxonomy} shows the second and third attack scenarios.

For example, modern 
anti-virus (AV) software uses DNNs to detect malware behaviour. 
Some AV modules detect such behaviours offline. 
When deploying a compressed model in such an application, 
how likely is it that malware could analyse the compressed model, work out how to evade the full model, and thus defeat the firm's other AV products?
Similarly, if an alarm company deploys a compressed model for 
intruder detection in consumer-grade CCTV equipment, could an intelligence agency
that buys such equipment figure out how to defeat not just that
product but government products derived from the same full model?
The risk is that just as a new type of software attack such as Heartbleed or Meltdown can cause widespread disruption by requiring thousands of disparate systems to be patched, so portable adverse examples could force upgrades to large numbers of diverse embedded systems.

\subsection{Networks and Compression Methods}
We use LeNet5 \cite{lecun2015lenet} and 
CifarNet \cite{zhao2018mayo} 
for our experiments on MNIST \cite{lecun2010mnist} and
CIFAR10 \cite{krizhevsky2014cifar} datasets.
The LeNet5 model has 431K parameters and classifies MNIST
digits with an accuracy of $99.36\%$.
The CifarNet classifier \cite{zhao2018mayo} has 1.3M parameters and 
achieves $85.93\%$ classification accuracy. 

We implemented two types of compression method: 
\begin{itemize}
    \item Fine-grained pruning on weights;
    \item Fixed-point quantisation of both weights and activations.
\end{itemize}

We used the Mayo tool to generate 
pruned and quantised models \cite{zhao2018mayo}, and
fine-tuned these models after pruning and quantisation.
For each pruning density or quantised bitwidth, 
we retrain $350$ epochs for LeNet5 and $300$ epochs for
CifarNet with three scheduled learning rate decays starting from $0.01$.
For each decay, the learning rate decreases by a factor of $10$.

Applying pruning on a pretrained model shrinks the number
of parameters and thus the memory
footprint of future AI ASICs.
We use fixed-point quantisation on
both weights and activations of a DNN.
Quantising both weights and activations means that
computations operate in low-precision fixed-point formats,
which cut the time and energy cost both data moves and computations.
For fixed-point quantisation, we use a 1-bit integer when bitwidth is 4,
a 2-bit integer when bitwidth is 8, and  4-bit integers for the rest of the fixed-point quantisations.
% In terms of adversarial attacks, quantization can be viewed as a form
% of adding noise to the network, and
% we are interested in how adding quantization noise in weights and activations
% would affect adversarial attacks. 

\subsection{Adversarial attacks}
In the work reported in this paper
we used three popular attacks developed in the research community. 
We now present mathematical definitions 
of the attacks and comments about their behaviour. 

% Basic Iterative Method 
% was developed by Kurakin et al \cite{DBLP:journals/corr/KurakinGB16} and 
% is basically an iterated version of the original fast gradient method (FGM) and 
% fast gradient sign method (FGSM) 
% proposed by .

Goodfellow et al. first introduced the
fast gradient method (FGM) and fast gradient sign method (FGSM) to
develop attacks~\cite{goodfellow2014explaining}. For the definitions 
we will use the following notation: 
$\boldsymbol{\theta}$ represents the parameters of the model, 
$\boldsymbol{X}$ represents the inputs, while
$y$ and $y_l$ represents the outputs and labels respectively.
We can then use
$J(\boldsymbol{\theta}$, $\boldsymbol{X}$, $y_{l})$ 
to represent the cost function. 
The original FGM and FGSM perturbations are computed as in 
\Cref{equ:fgm} and \Cref{equ:fgsm} respectively, 
where $\epsilon$ is a hyperparameter and the function $\nabla_{X}()$
computes the first-order derivative with respect to input $X$.

\begin{equation}
\eta = \epsilon(\nabla_{X} J(\boldsymbol{\theta}, \boldsymbol{X},y)) 
\label{equ:fgm}
\end{equation}

\begin{equation}
\eta = \epsilon\text{sign}(\nabla_{X} J(\boldsymbol{\theta}, \boldsymbol{X},y)) 
\label{equ:fgsm}
\end{equation}

\begin{figure*}[!ht]
\centering
\begin{minipage}{.33\linewidth}
  \centering
  \includegraphics[width=\linewidth]{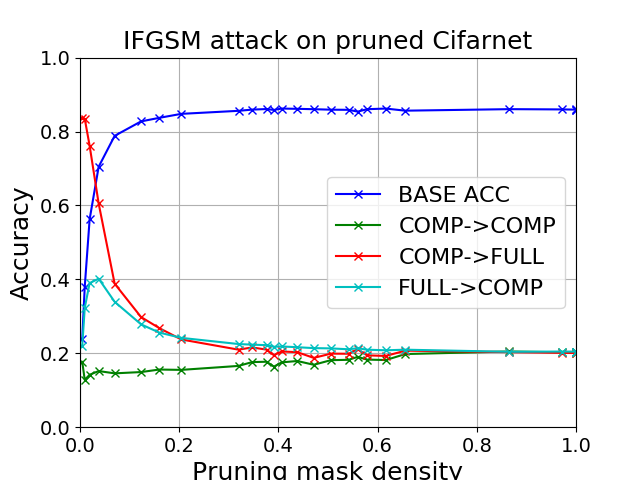}  
  \includegraphics[width=\linewidth]{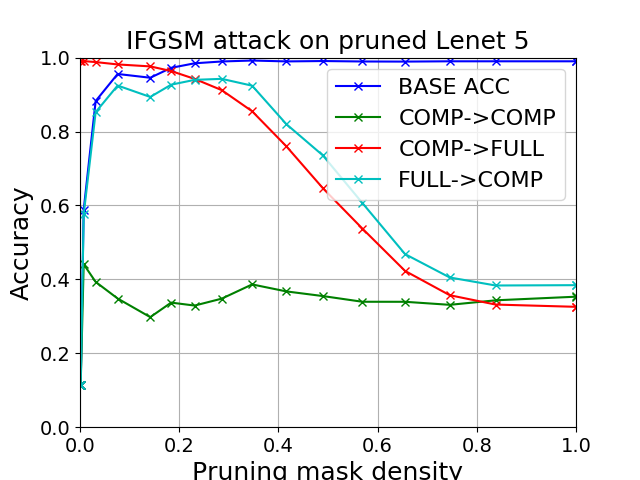}
\end{minipage}
\begin{minipage}{.33\linewidth}
  \centering
  \includegraphics[width=\linewidth]{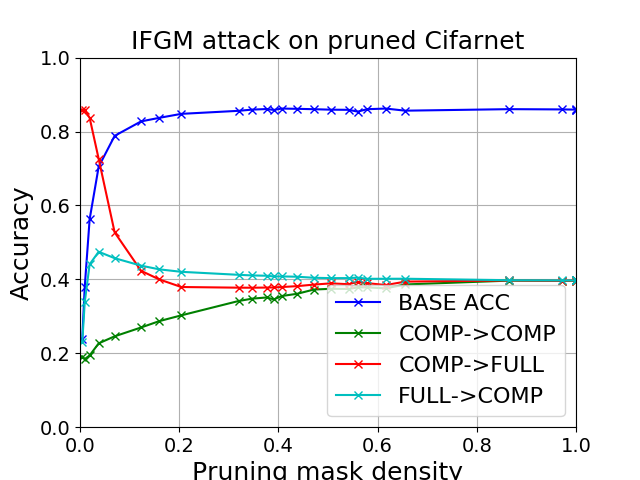}  
  \includegraphics[width=\linewidth]{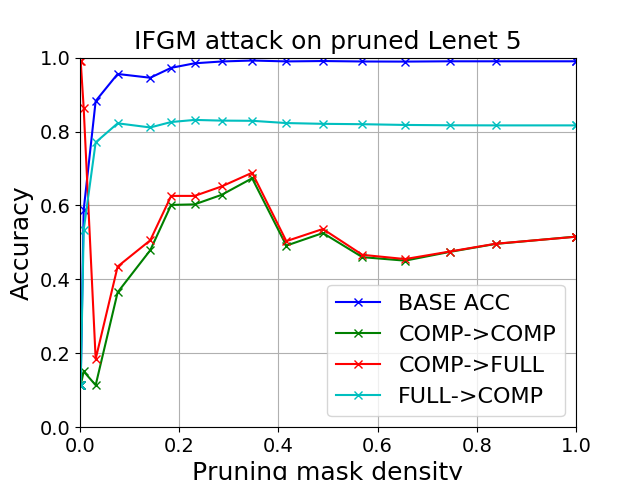}
\end{minipage}
\begin{minipage}{.33\linewidth}
  \centering
  \includegraphics[width=\linewidth]{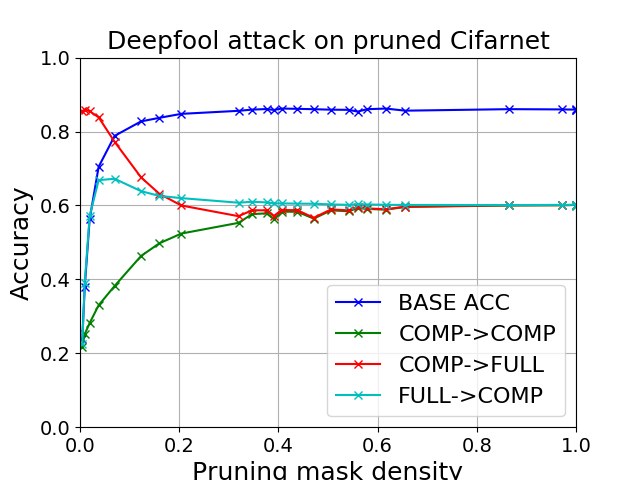}  
  \includegraphics[width=\linewidth]{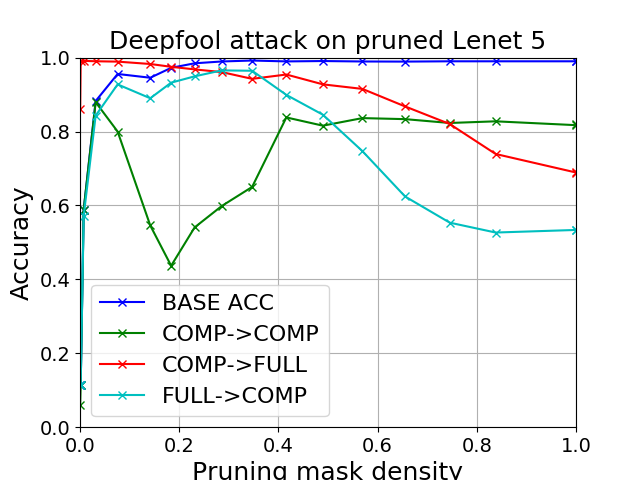}
\end{minipage}
\caption{Transferability properties for pruning.
The green, red and cyan lines represent the first, second and third
attack scenarios respectively. The blue line show the accuracies of pruned
models without any attacks.}
\label{fig:pruning_transferability}
\end{figure*}

Kurakin et al. presented an iterative algorithm based on FGM and FGSM methods~\cite{DBLP:journals/corr/KurakinGB16}.
In \Cref{alg:fgsm},
we present an iterative FGSM (IFGSM),
where the adversarial samples $X_n^{\text{adv}}$ are generated
for the $n$th iteration.
% The iterative FGM (IFGM) is nearly
% identical to the IFGSM with a change of 
% $\boldsymbol{N} =  \text{sign}(\nabla_{X}J(\boldsymbol{\theta}, \boldsymbol{X}_{n}^{\text{adv}},y_{l}))$.

During each iteration, 
the intermediate results get clipped 
to ensure that the resulting adversarial images
lie within $\epsilon$ of the previous iteration. 

\begin{algorithm}[h]
   \caption{IFGSM}
   \label{alg:fgsm}
\begin{algorithmic}
   \STATE {\bfseries Input:} data $X_{in}$
   \STATE Initialize $X_0^{adv} = X_{in}$.
   \FOR{$n=0$ {\bfseries to} $m-1$}
   \STATE $\boldsymbol{N} =  \epsilon\text{sign}(\nabla_{X}J(\boldsymbol{\theta}, \boldsymbol{X}_{n}^{\text{adv}},y_{l}))\}$
   \STATE $\boldsymbol{X}_{n+1}^{\text{adv}} = Clip_{X,\epsilon}\{\boldsymbol{X}_{n}^{\text{adv}} + \boldsymbol{N}\}$
   \ENDFOR
\end{algorithmic}
\end{algorithm}

Kurakin et al.~also presented 
an iterative version of FGM 
where instead of just using the 
sign to determine the direction of a gradient, 
the gradient amplitudes contribute to 
the gradient update step. 
The iterative FGM (IFGM) is nearly
identical to the IFGSM except that
$\boldsymbol{N} = \epsilon\nabla_{X}J(\boldsymbol{\theta}, \boldsymbol{X}_{n}^{\text{adv}},y_{l})$.

Moosavi-Dezfooli et al. featured another attack called `Deepfool', which is also based on 
iterative gradient adjustment~\cite{DBLP:journals/corr/Moosavi-Dezfooli15}.
However, Deepfool is different from IFGSM in that 
it does not scale and clip gradients. 
It is based on the idea that the
separating hyperplanes in linear classifiers indicate the 
decision boundaries of different classes.
It therefore iteratively perturbs an image $\boldsymbol{X}_{0}^{adv}$, 
linearises the classification space around $\boldsymbol{X}_{n}^{adv}$
and moves towards the closest decision boundary. The step 
is chosen according to the $l^{0}$, $l^{1}$ or even the $l^{p}$ norm of 
$\boldsymbol{X}_{n}^{adv}$ to the last-found decision boundary. 
The applied step is then used as $\boldsymbol{X}_{n+1}^{adv}$. 

In practice Deepfool is found to produce 
smaller perturbations than the original IFGSM, which makes 
it a more precise attack \cite{DBLP:journals/corr/Moosavi-Dezfooli15}.
In this paper we used an L2 norm-based version of Deepfool.

It should be noted that in this particular paper we were not interested in
the absolute accuracy but the relative behaviours with a set of fixed
parameters for adversarial attacks.
We chose the strongest white box adversary model and picked three of the strongest iterative attacks.
For all the experiments, we did not sweep all the possible hyper-parameters for the adversarial attacks, but picked empirically sensible hyper-parameters. The parameters are shown in~\Cref{tab:attack_hyperparam} and were chosen in such a way that they generated perturbations of a sensible $l^2$ and $l^0$ and caused noticeable classification change. 

Finally, we want to talk about how realistic the scenarios presented in this paper are and why we chose those particular attacks. First, we want to mention that not all of the attacks are transferable -- as a matter of fact the attacks we chose are amongst the least transferable ones. We chose those specific attacks to explore the lower bound of transferability and show how much of the subspaces actually survive the compression process~\cite{tramer2017space}. For DeepFool, we trained two models with different random initialisation and tested how tranferable the adversarial samples are. For LeNet5 only 7\% of the samples actually went across, whereas for CifarNet the transferability was better, but still only 60\%. We now describe a slightly different scenario -- the models attacking and defending are the same, just some of them are compressed. 

\begin{table*}
\centering
\begin{tabular}{@{}c|cc|cc|cc@{}}
\toprule
\multirow{2}{*}{Network/Attack} & \multicolumn{2}{|c}{I-FGSM} & \multicolumn{2}{|c}{I-FGM} & \multicolumn{2}{|c}{DeepFool} \\ 
& $\epsilon$ & $i$ & $\epsilon$ & $i$ & $\epsilon$ & $i$ \\ \midrule
LeNet5         & 0.02 & 12 & 10.0 & 5 & 0.01 & 5 \\
CifarNet       & 0.02 & 12 & 0.02 & 12 & 0.01 & 3     \\ \bottomrule
\end{tabular}
\caption{Attack hyper-parameters.}
\label{tab:attack_hyperparam}
\end{table*}

For all of the attacks we made sure that in each of the iterations the perturbations stayed within the expected range.

% Papernot et al. \cite{DBLP:journals/corr/PapernotMJFCS15}
% presented `Jacobian-based Saliency Map Attack' (JSMA) 
% a new way to generate the adversarial samples. 
% Unlike the attacks presented previously,
% the attack relies on forward derivative computation on the input itself, 
% rather then the cost function. 
% From a calculated derivative the saliency map gets built and 
% the inputs get changed according to it. 
% This gets repeated until the the class changes successfully or
% the pixel change reaches its maximum.

% Carlini and Wagner \cite{Carlini017}
% have presented three attacks (CW) formed as an optimisation problem. 
% The attacks themselves differ by the p-norm order used in the 
% objective function according to which the optimisation takes place. 
% In order to get the best results one needs to fine-tune 
% a range of hyperparameters. 
% For our paper, we have used CW based on the L2 norm 
% and a temperature equal to 2 and $\kappa=2$. 
% \todo{what is k, seems undefined}

\section{Evaluation}

\subsection{Pruning}

\begin{figure*}[!ht]
\centering
\begin{minipage}{0.45\linewidth}
  \centering
  \includegraphics[width=\linewidth]{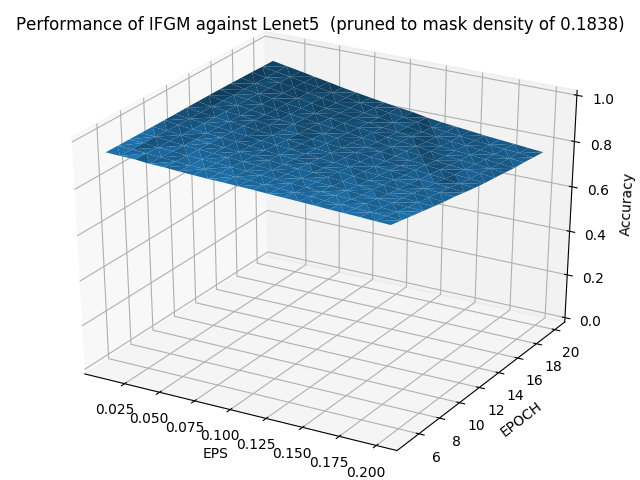}
\end{minipage}
\begin{minipage}{0.45\linewidth}
  \includegraphics[width=\linewidth]{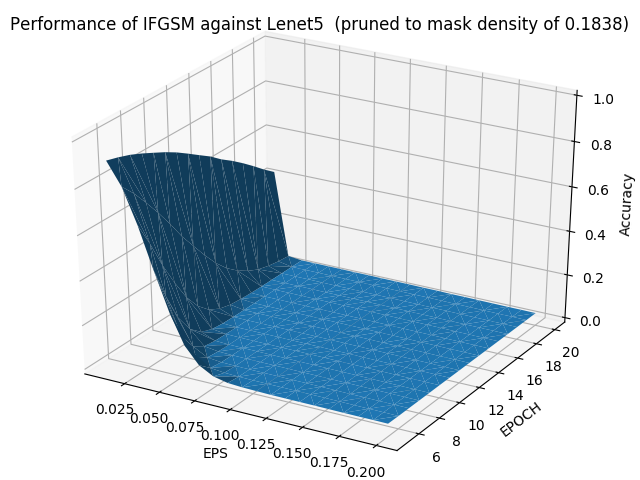}
\end{minipage}
\caption{Lenet5 accuracy with IFGSM and IFGM-generated adversarial samples with different $epsilon$ values and number of epochs}
\label{fig:exp2_lenet_adv_accuracies-pruned_13_fgsm}
\end{figure*}

Goodfellow et al. explained the existence of adversarial samples
as follows \cite{goodfellow2014explaining}.
Consider an adversarial sample as the original input $\boldsymbol{x}$
with an additional noise $\boldsymbol{\eta}$.
When passing through multiple layers of matrix multiplication, this
small noise eventually grows to a large enough value
to shift the decision of the whole model.
Given weights $\boldsymbol{w}$ of a particular layer of a
neural network and adversarial sample
$\widetilde{\boldsymbol{x}}=\boldsymbol{x}+\boldsymbol{\eta}$,
the output of that particular layer is
$\boldsymbol{w}^{\boldsymbol{T}}\widetilde{\boldsymbol{x}}=\boldsymbol{w}^{\boldsymbol{T}}\boldsymbol{x}+\boldsymbol{w}^{\boldsymbol{	T}}\boldsymbol{\eta}$,
and the adversarial perturbation causes the the output activations to grow by
$\boldsymbol{w}^{\boldsymbol{	T}}\boldsymbol{\eta}$.

\Cref{fig:pruning_transferability} shows the performance of IFGSM, IFGM and DeepFool
on pruned models under three different attack scenarios.
The horizontal axis shows the densities of DNNs, effectively the
ratio of the number of non-zero values to the total number of values.
The vertical axis presents test accuracies of DNNs.
Apart from showing the accuracies of pruned networks without any attacks
(BASE ACC), we present the accuracies of the pruned models with
three different attack scenarios.
The first scenario corresponds to COMP $\to$ COMP,
the second scenario and third scenario corresponds to
FULL $\to$ COMP and COMP $\to$ FULL respectively.

The first thing to note is that samples generated
from the compressed models
are transferable to the baseline model when densities are relatively large.
This finding reinforces the idea that the adversarial samples
are not scattered randomly but reside in
large and contiguous high-dimensional spaces, enabling them to survive the
effects of pruning.
We suggest that pruning smooths the decision space
by removing DNN weights that have little impact.
This ultimately has an effect on IFGSM
-- with unimportant parts removed, the gradients now follow the path
towards the most important and prominent parts of the space (first and fourth plots on \Cref{fig:pruning_transferability}).
As a result,
relatively small perturbations based on compressed models generalise very well
on the uncompressed model when networks are not heavily pruned.
For all of the attacks,
adversarial samples generated on networks with
very small densities are not effective on the baseline networks
(increase in the red line and fall of the blue line near zero in all plots of \Cref{fig:pruning_transferability})
Heavily-pruned networks acquire a feature space
that is hugely different from the baseline models, and this
limits the transfer of adversarial samples.
However, low-density networks often suffer large losses in classification
accuracy, making them infeasible to deploy in real life.

When one compressed model is attacking another (Comp to Comp, green line on Figure 2),
we see a general trend that attacks remain transferable.

% the decrease in attack efficacy should be attributed to the fact that
% the previously-used step $\epsilon$ was no longer large enough to reach a
% successful decision boundary along the gradients.
% \Cref{fig:exp2_lenet_adv_accuracies-pruned_13_fgsm} shows that small
% $\epsilon$ values are not efficient for attacking pruned models.
% On both IFGM and IFGSM, we notice an increase in accuracy when
% the $\epsilon$ values are small.
% Also, IFGSM shows a better attacking performance but suffers more
% from the effect of small $\epsilon$ values.
% Such behaviour in case of IFGM and IFGSM can be
% explained by the use of the \textit{Clip} function.
% The use of $\epsilon$ and \textit{Clip} function was originally designed
% to make sure that the changes of the original images are not too large,
% and apparently the attacks require a large $\epsilon$ to remain
% effective on pruned models.

\begin{figure*}[!ht]
\centering
\begin{minipage}{0.45\linewidth}
  \includegraphics[width=\linewidth]{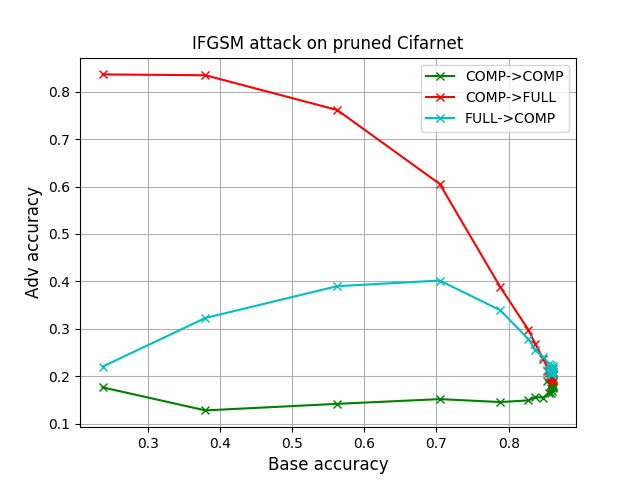}
\end{minipage}
\begin{minipage}{0.45\linewidth}
  \centering
  \includegraphics[width=\linewidth]{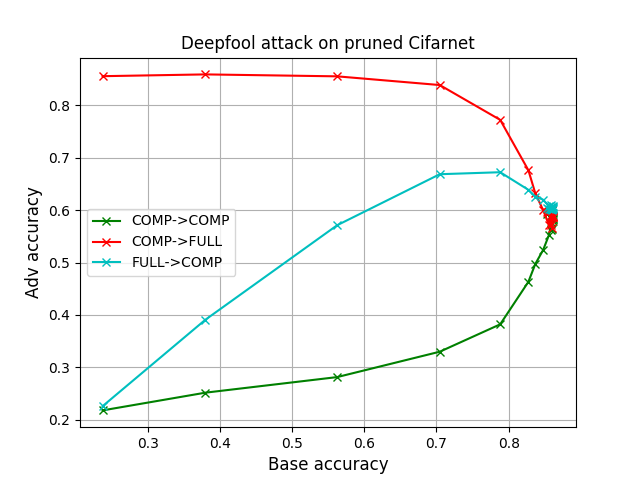}
\end{minipage}
\caption{CifarNet accuracy with IFGSM and DeepFool-generated adversarial samples with corresponding base accuracy.}
\label{fig:pruning_pareto}
\end{figure*}

Using an uncompressed model to attack a compressed
one (cyan line, \Cref{fig:pruning_transferability}), we observe
a slight increase in accuracy occurs when densities
are small, but then a rapid drop when they keep decreasing;
this effect occurs when the base accuracy (blue line) also starts to drop.
We view pruning as a regularization method which removes local
minima from the large optimization space.
When the blue line is just starting to decrease, this turning point is the preferred density, where the network just stops overfitting.
In other words, the preferred density of a network represents the minimal number of parameters needed to accomplish the same accuracy
as the dense model.
To further illustrate the effect of pruning, we present CifarNet with
Deepfool and IFGSM in \Cref{fig:pruning_pareto}.
The horizontal axis shows the accuracy of the baseline model and
the vertical axis has the adversarial accuracy.
In terms of adversarial attacks, although limited, we observe that networks that reach preferred density show a protective nature in \Cref{fig:pruning_pareto}.

We observe attacks perform worse on LeNet5 in comparison to CifarNet.
We notice that LeNet5 inherently achieves a larger accuracy on MNIST,
meaning that the loss is smaller on the evaluation dataset as well.
When building attacks, we often make use of the amplitude of gradients to generate adversarial samples, and the smaller loss associated with LeNet5 implies that it is less vulnerable to attack.
This phenomenon becomes more apparent on Deepfool and IFGM, since these attacks employ the gradients amplitudes to generate adversarial images.

Intuitively,
pruning largely preserves the feature space of a baseline CNN,
so adversarial samples remain transferable.
This empirical observation confirms recent pruning discoveries, that pruned networks
distil feature spaces~\cite{liu2018rethinking,frankle2018lottery}.
In addition,
our observation is in line with the suggestions made
by Tramer et al.:
if the feature spaces are similar, adversarial samples stay highly transferable~\cite{tramer2017space}.

%Under the first and second scenarios,
%we observe FGM and FGSM are less effective when networks are at
%lower densities.
%In contrast, DeepFool remains to be an effective attacking method
%when networks are sparse.
%Intuitively, iterative FGM and FGSM contains clipping
%and taking signs of the gradients.
%The amount of noise added to the samples are more coarse-grained.
%In contrast, DeepFool specializes in tuning the magnitudes of
%noises added that causes a misclassification.
%Recall that during pruning a mask consisting of zeros and
%ones gets applied to the weights.
%The expected effect of pruning thus is
%that the adversarial changes now get zeroes out and
%to find an adversarial sample becomes a more fine-grained problem.
%One intuitive explanation is that, pruning serves as a regularization method
%and helps networks to generalize better.
%To attack a network that has better generalizations, a more fine-grained
%attacking method is required for creating adversarial samples.
%DeepFoll, therefore, shows better attacking performance compared to
%FGM and FGSM.

%In addition, we observe that, in scenario 3,
%adversarial samples generated by pruned models can be very
%effective on the baseline
%model.
%Intuitively, generating adversarial samples on a pruned model
%is to generate samples that worked well on neural networs that generalize well.
%If such a harder attack successed, the same adversarial attack can be easily
%applied on the baseline model.

% \subsubsection{Summary}
\textbf{Summary}
\begin{enumerate}
\item The transferability of adversarial samples
 between pruned and full models remains when networks are slightly pruned.
 \item For compressed models attacking uncompressed models, we observe
 worse transferability when models are heavily pruned, but the
 original accuracy of these sparse models decreases significantly.
 \item For uncompressed models attacking compressed models,
 the accuracy is maximised, and the transferability minimised, at the preferred density where the network stops overfitting.
\end{enumerate}

\begin{figure*}[!ht]
\centering
\begin{minipage}{.33\linewidth}
  \centering
  \includegraphics[width=\linewidth]{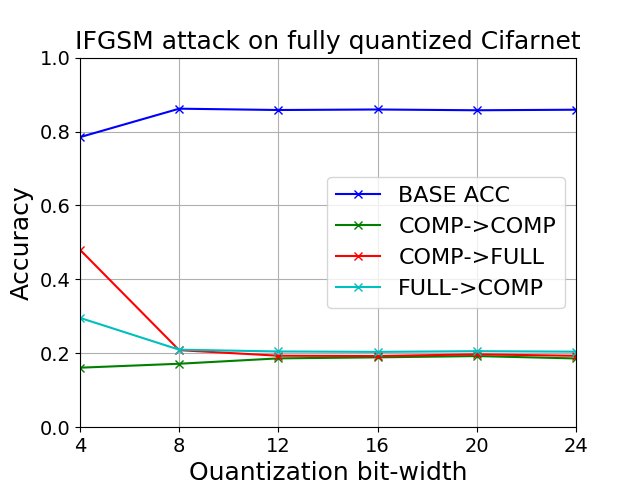}
  \includegraphics[width=\linewidth]{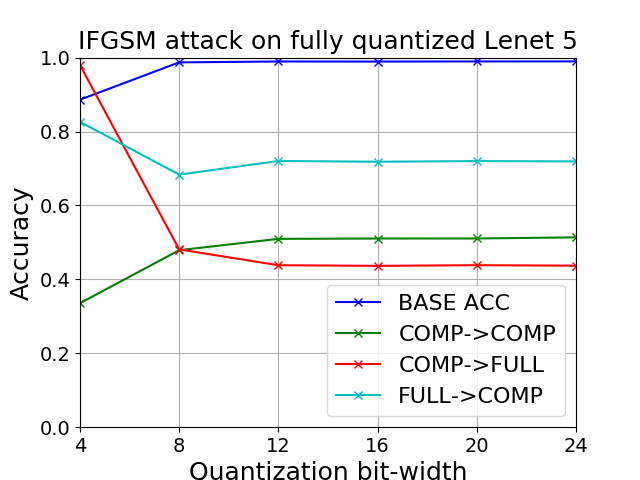}
\end{minipage}
\begin{minipage}{.33\linewidth}
  \centering
  \includegraphics[width=\linewidth]{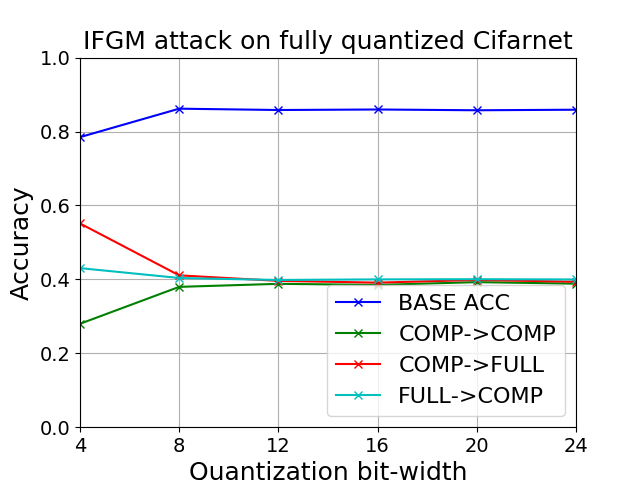}
  \includegraphics[width=\linewidth]{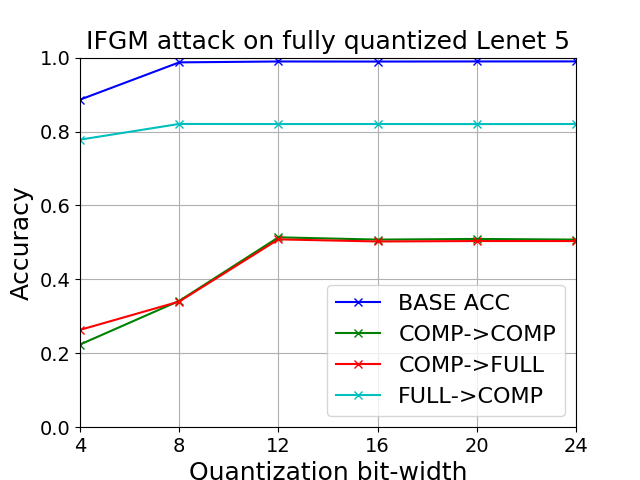}
\end{minipage}
\begin{minipage}{.33\linewidth}
  \centering
  \includegraphics[width=\linewidth]{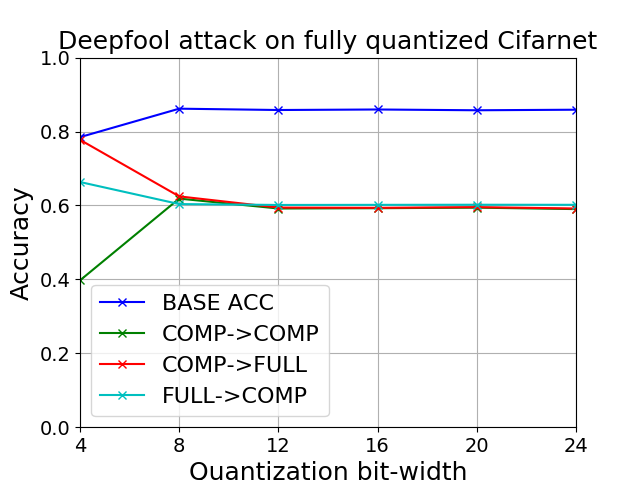}
  \includegraphics[width=\linewidth]{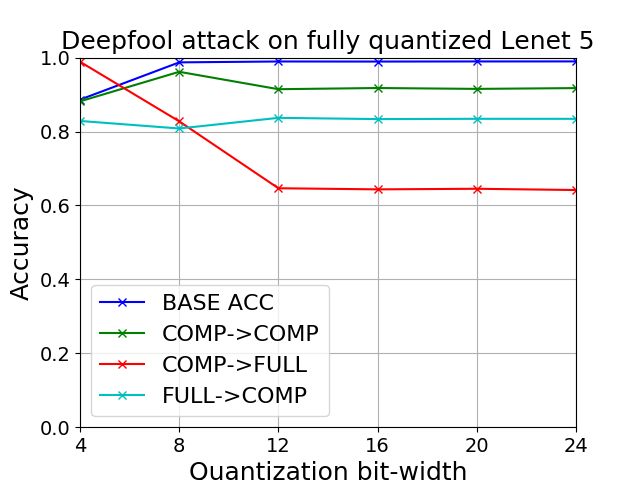}
\end{minipage}
\caption{Transferability properties for quantising both weights and activations.
The green, red and cyan lines represent the first, second and third attack scenarios respectively. The blue line show the accuracy of quantised models without any attacks.}
\label{fig:quan_transferability}
\end{figure*}

% \begin{figure*}[!ht]
% \centering
% \begin{minipage}{.33\linewidth}
%   \centering
%   \includegraphics[width=\linewidth]{images/final_cifar_quan_fgsm_combined_adv_accuracies_1_31_32.png}
%   \includegraphics[width=\linewidth]{images/final_lenet_quan_fgsm_combined_adv_accuracies_1_31_32.png}
% \end{minipage}
% \begin{minipage}{.33\linewidth}
%   \centering
%   \includegraphics[width=\linewidth]{images/final_cifar_quan_fgm_combined_adv_accuracies_1_31_32.png}
%   \includegraphics[width=\linewidth]{images/final_lenet_quan_fgm_combined_adv_accuracies_1_31_32.png}
% \end{minipage}
% \begin{minipage}{.33\linewidth}
%   \centering
%   \includegraphics[width=\linewidth]{images/final_cifar_quan_deepfool_combined_adv_accuracies_1_31_32.png}
%   \includegraphics[width=\linewidth]{images/final_lenet_quan_deepfool_combined_adv_accuracies_1_31_32.png}
% \end{minipage}
% \caption{Transferability properties for weights-only quantisation}
% \label{fig:quan_transferability}
% \end{figure*}

\subsection{Fixed-point Quantisation}

Fixed-point quantisation refers to quantising
both weights and activations
to fixed-point numbers; for example, 
for a bitwidth of four we use a one-bit integer plus a three-bit fraction.
Multiplication is much faster as we can use integer operations rather than floating point. \Cref{fig:quan_transferability} shows the performance of adversarial
attacks on quantised models under our three different attack scenarios.
Attack performance stays nearly constant
at bitwidths higher than 8.
When using fewer bits for both weights and activations,
the model shows defensive behaviour, mainly because of the reduction in
integer precision.

\begin{figure*}[!ht]
\centering
\begin{minipage}{0.5\textwidth}
  \centering
  \includegraphics[width=\linewidth]{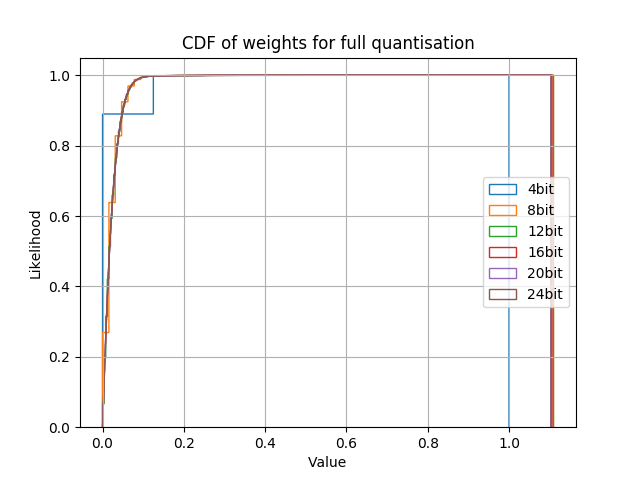}\\
  (a) Weights
\end{minipage}%
\begin{minipage}{.5\textwidth}
  \centering
  \includegraphics[width=\linewidth]{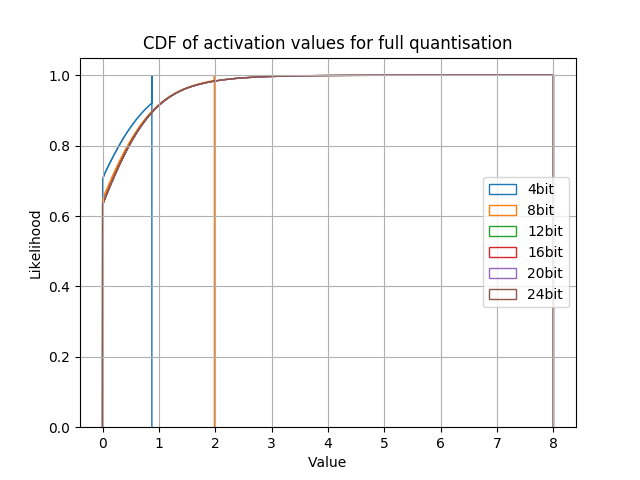}
  (b) Activations
\end{minipage}
\caption{Cumulative distribution function (CDF)
for all weights and activations in quantised CifarNet. Ten randomly chosen input images from the validation dataset were used for generate CDF of activation values.}
\label{fig:w_act_dist}
\end{figure*}

Intuitively, we have two effects when
values are quantised to smaller bitwidths.
First, a smaller bitwidth means fewer fractional bits causing a loss in precision, and
introducing much the same effect as pruning.
Second, it can mean fewer integer bits,
so weight and activation values are smaller.
Thus, our models in 4-bit
fixed-point quantisation have smaller
weight and activation values
and contain more zeros than models at higher precisions.
In \Cref{fig:w_act_dist}.a, we show the cumulative distribution function (CDF)
of CifarNet with different fixed-point quantisations.
There are clearly more zeros in the 4-bit CifarNet --
its cumulative density reaches around $0.9$ when value is at $0$.
The clipping effect is also more obvious in the 4-bit model, since it
only has a 1-bit integer part; we can see the 4-bit model has its weights CDF reach
1.0 before all other bitwidths in \Cref{fig:w_act_dist}.a.

Using the adversarial examples generated by compressed models
to attack the baseline model (red line),
we observe both Deepfool and IFGSM
methods become less effective on LeNet5 and CifarNet.
The same phenomenon occurs when we use adversarial samples
generated by the baseline model to attack quantised models (cyan line).
We suggest that during quantisation, reducing fractional bits will not
hugely impact the attacks' performance at high bitwidths,
but introduces a similar effect to pruning at low bitwidths.
In addition,
reducing integer bits essentially
introduces large differences between the baseline model and
the quantised ones for adversarial attacks.
Using FGM to attack LeNet5, on the other hand, gives very different behavior.
We also noticed that attacking LeNet5 require large epsilon values
and more iterative runs.
We suggest this is because of the accuracy issue with LeNet5 that
we've addressed earlier -- attacks that rely on gradient magnitudes 
struggle with networks that achieve high accuracy.

By reducing the length of the fraction, the rounding
process of fixed-point quantisation becomes more lossy.
Uniformly adding quantisation noise to each individual weight
does not affect attack performance.
As we can see in \Cref{fig:quan_transferability}, all three attack scenarios
show a stable performance when bitwidths are higher than 8,
where the difference
lies in the length of the fraction.
When the network gets quantised down to 4 bits,
quantisation behaves rather like pruning --
a large part of the network gets zeroed out.

By reducing integer bitwidth,
we are clipping the numerical values.
Theoretically, clipping weights is different from
clipping activations.
For the former,
we first consider how to create an adversarial sample with minimal perturbation.
Intuitively, the way to achieve such a perturbation with minimal changes
on the input image is tweak
pixels with large weight values that are connected
to important activations.
Thus a small change in input image pixels can
have the maximal effect on activation values.
When weights are clipped,
adversarial attacks see more weights as having equal importance because
they saturate to the same maximal value.
This undermines attack transferability between quantised models
and baseline models.
For example, on a quantised network, an adversarial example $X_i$
considers $w_i = \max(W_i)$ to be the largest weight associated with the important
activations among all the weights ($W_i$) associated with activation $a_i$.
This relationship $w_i = \max(W_i)$ might break on the baseline model and thus
the adversarial sample becomes less effective.
In \Cref{fig:w_act_dist}, a 4-bit fixed-point quantisation clearly
shows a clipping effect on weight values, which contributes to
the marginal defensive nature we observed in \Cref{fig:quan_transferability}.

When activations are clipped to a smaller maximal value,
transferability between quantised
and baseline models becomes worse.
\Cref{fig:w_act_dist}.b shows how activations are clipped to different
maximum values.
Consider a simple case, where an adversarial example overdrives
one activation to be larger than others
in the same layer to cause a misclassification.
%When activations are clipped,
%this over-driving now can maximally push values to the clipped value.
Clipping the activation values forces the attacker to find more subtle  
ways of achieving differential activation, which
is significantly harder.

Clipping weights and clipping activations can both significantly
affect attack performance.
As we can see from the cyan line in \Cref{fig:quan_transferability}, at smaller bit widths, all our attack scenarios show
an increase in accuracy. except for LeNet5 attacked by IFGM.
In terms of transferability, adversarial examples remain transferable between quantised and baseline models under both IFGSM and IFGM 
when fractional bits are lost, but integer bits start to decrease, transferability
becomes worse.

%Saturation happens when the integer bitwidth is small, as big values are capped.
%In other words, weights of a quantised DNN are clustered into a smaller groupof values that are closer together.
%A small perturbation can nowaffect a large number of weights since many weightsnow have similar importance.
%\Cref{fig:final_cifar_act_quan_fgsm_combined_mean_pertubations_1_31_32} shows how adversarial samples generated from quantised DNNsnow require smaller perturbations to attack the baseline model.
% However, the smaller bitwidths have a negative effect
% on transferability of the generated samples:.
% For scenario 2, small changes might be within the boundaries of baseline model.
% For scenario 3, big changes might be
% outside of the boundaries of being distinguishable for a compressed model.
% We attribute the slight decrease in IFGM and IFGSM performance
% at 4-bit to these reasons.

Surprisingly, we find that Deepfool, unlike IFGM and IFGSM,
struggles to generate effective adversarial samples when models are quantized.
%We attribute this to the discrete levels introduced by quantisation.
Restricting values to discrete levels mean now an attack has to
inject a large enough perturbation to the values to push them to the neighbouring
quantizaton level.
Since Deepfool is a very fine-grained attack, it struggles to do this.

Although IFGSM shows slightly higher accuracy at low precision,
this protective behavior is only marginal.
Attacks still show good performance
in all three scenarios compared to both IFGM and DeepFool if we consider the
classification accuracy (\Cref{fig:quan_transferability}).

In summary, transferability is still a hazard for quantized models.
But there is some marginal protection when activation values are quantized.
Intuitively, this forces many activations to saturate and thus makes it harder
for the attacker to overdrive certain values.

\textbf{Summary}
\begin{enumerate}
 \item The transferability of adversarial samples
 between quantised and baseline models is not affected by
 reducing fractional bitwidth at high precision.
 \item Aggressive reduction of fractional bits introduces the same
 effect as fine-grained pruning.
 \item Smaller integer bitwidths of weights and activations
 make it marginally harder to attack
 the baseline model using adversarial samples generated from  compressed models.
 This suggests that the network's knowledge is contained
 in both activations and weights.
 \item We hypothesize that clipping activations changes the feature space of CNNs and thus marginally protects models from transferability.
\end{enumerate}

\section{Conclusion}
This paper reports an empirical study of the interaction 
between adversarial attacks and neural network compression.

Both quantisation and pruning sparsify the network, i.e. a
greater number of activiations and weights are zero. 
Attacks generated from heavily pruned models
work effectively against the underlying baseline model.
However, low-density DNNs are somewhat defensive when attacked by
adversarial samples generated from the baseline model using fast-gradient-based methods.
Quantisation is different in that adversarial samples 
from fast-gradient-based methods become marginally
harder to transfer when models are heavily quantised.
This defensive behaiour appears due to the reduction in integer bits 
of both weights and activations rather than to the truncation in fractional bits.

The broader implications are that attacks on DNN classifiers that involve adversarial inputs may be surprisingly portable. Even if a
firm ships only a compressed version of its classifier in widely
distributed products, such as IoT devices or apps, attacks that people discover on these compressed classifiers may translate fairly easily to attacks on the underlying baseline model, and thus to other compressed versions of the same model. Just as software vulnerabilities such as Heartbleed and Spectre required the patching of many disparate systems, so also a new adversarial sample may defeat many classifiers of the same heritage. Firms should be aware that while shipping a compressed classifier may give real benefits in terms of performance, it may not provide much in the way of additional safety or security.

\section*{Acknowledgements}
\textit{Partially supported with funds from Bosch-Forschungsstiftung im Stifterverband.}

\bibliographystyle{unsrt}
\bibliography{references.bib}

\end{document}